\documentclass{natureTEST}

\usepackage{graphicx}
\usepackage{bm}
\usepackage{amsmath,amssymb}
\usepackage{xspace}
\usepackage[rightcaption]{sidecap}

\usepackage{textcomp}
\usepackage[T1]{fontenc} 

\usepackage[switch]{lineno} 

\usepackage[colorlinks=false, pdfborder={0 0 0}]{hyperref}

\newcommand{\be}{\begin{eqnarray}}
\newcommand{\ee}{\end{eqnarray}}

\newcommand{\edd}{\epsilon_{\rm dd}}

\title{Self-bound droplets of a dilute magnetic quantum liquid}

\author{Matthias Schmitt$^{1}$, Matthias Wenzel$^{1}$, Fabian B\"ottcher$^{1}$, Igor Ferrier-Barbut$^{1}$ \& Tilman Pfau$^{1}$}

\begin{document}

\maketitle

\begin{affiliations}
	\item 5. Physikalisches Institut and Center for Integrated Quantum Science and Technology, Universit\"at Stuttgart, Pfaffenwaldring 57, 70550 Stuttgart, Germany
\end{affiliations}

\begin{abstract}
Self-bound many-body systems are formed through a balance of attractive and repulsive forces and occur in many physical scenarios. Liquid droplets are an example of a self-bound system, formed by a balance of the mutual attractive and repulsive forces that derive from different components of the inter-particle potential. It has been suggested\cite{Petrov:2015,Bulgac:2002} that self-bound ensembles of ultracold atoms should exist for atom number densities that are 10$^\mathbf{8}$ times lower than in a helium droplet, which is formed from a dense quantum liquid. However, such ensembles have been elusive up to now because they require forces other than the usual zero-range contact interaction, which is either attractive or repulsive but never both. On the basis of the recent finding that an unstable bosonic dipolar gas can be stabilized by a repulsive many-body term\cite{Ferrier:2016}, it was predicted that three-dimensional self-bound quantum droplets of magnetic atoms should exist\cite{Waechtler:2016a,Baillie:2016}. Here we report the observation of such droplets in a trap-free levitation field. We find that this dilute magnetic quantum liquid requires a minimum, critical number of atoms, below which the liquid evaporates into an expanding gas as a result of the quantum pressure of the individual constituents. Consequently, around this critical atom number we observe an interaction-driven phase transition between a gas and a self-bound liquid in the quantum degenerate regime with ultracold atoms. These droplets are the dilute counterpart of strongly correlated self-bound systems such as atomic nuclei\cite{Bender:2003} and helium droplets\cite{Dalfovo:2001}. 
\end{abstract}

Liquid droplets of water or helium are formed by the mutual attractive and repulsive forces that are created by the different parts of the inter-particle potential (and are due to covalent or van der Waals attraction and to the electronic Pauli exclusion principle, respectively). Helium droplets in particular have been a focus of research, owing to their interesting quantum nature\cite{Volovik:2009,Toennies:2004}. Droplets can serve as closed, isolated quantum systems with which to probe, for example, superfluidity of mesoscopic ensembles\cite{Gomez:2014}. In the context of ultracold atoms, the observation of an ensemble of stable droplets\cite{Kadau:2016} in a dilute magnetic quantum gas opened up the possibility of a three-dimensional self-bound state\cite{Waechtler:2016a,Baillie:2016}. A trapped quantum droplet of magnetic atoms has recently also been observed using erbium atoms\cite{Chomaz:2016}. Here we demonstrate the observation of dilute, self-bound liquid droplets in a sample of ultracold bosonic dysprosium atoms, which have a strong longrange magnetic dipolar interaction and a tunable repulsive short-range contact interaction. The interplay between these two interactions can be tuned such that the overall mean field is weakly attractive, but so that the interactions also create quantum depletion and a corresponding many-body repulsion. This repulsion exactly counteracts the attraction when the density of the droplet reaches the stabilization density. We use the word 'liquid' here to describe a state of matter that is defined by the presence of self-bound droplets and by the stabilization of the self-binding forces as a result of repulsion beyond the simple meanfield level, which manifests itself as a nontrivial correlation function. For dilute liquids, these correlations can be very weak (as in the present case), contrary to dense liquids for which correlations are strong. At small atom numbers (around 1,000 atoms), the finite size of the wavefunction of the quantum droplet leads to a quantum pressure for each individual atom that results in an evaporation out of the self-binding potential. Therefore, these droplets are bound only above a critical atom number, which we investigate systematically.

\begin{figure}[!t]
\begin{center}
\includegraphics[width=\columnwidth]{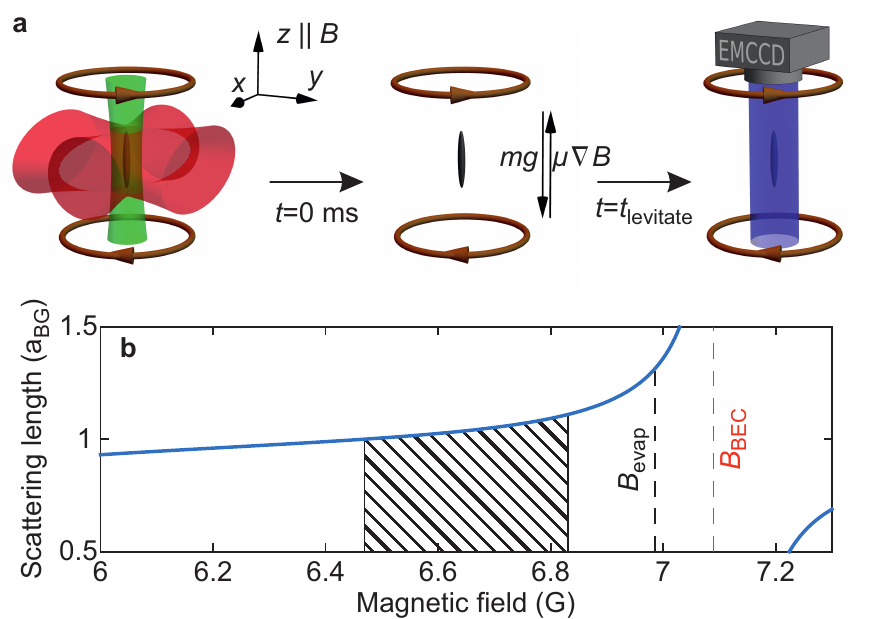}\\
\end{center}
\caption{\textbf{Experimental sequence. a}, A schematic of the experimental sequence. We start with an atomic ensemble in a crossed optical trap superimposed with a magnetic field gradient that is strong enough to compensate the gravitational force. We then turn off the trapping beams and levitate the droplet for various times $t_{\rm levitate}$. Finally, we image the atoms using phase-contrast polarization imaging projected on an EMCCD (electron multiplying charge coupled device) camera. \textbf{b} Scattering length as a function of magnetic field at the region of the Feshbach resonance in units of the positive background scattering length $a_{\rm BG}$. The red dashed
line indicates the field ($B_\mathrm{BEC}$) at which we create a BEC. The hatched area describes the region in which the experiments were performed. The dashed black line shows the magnetic field ($B_\mathrm{evap}$) used to intentionally evaporate the droplets to the gas phase.}
\label{Fig:FBscheme}
\end{figure}

We use \textsuperscript{164}Dy, which has one of the strongest magnetic dipole moments in the periodic table with $\mu = 9.93\,\mu_B$, where $\mu_B$ is the Bohr magneton. These atoms also offer control on the short-range interaction by a magnetic field using Feshbach resonances\cite{Chin:2010,Tang:2015,Maier:2015b}. Here we use a specific resonance at a field of $B_0=7.117(3)$\,G with a width of $\Delta B=51(15)\,$mG. (Here and elsewhere, the errors in parentheses indicate one standard deviation.) This resonance allows the scattering length a to be tuned from that of a dipole-dominated sample to a contact-dominated sample, without strong losses (Fig. \ref{Fig:FBscheme}b). To quantify the relative influence of the short-range and dipole-dipole interactions, we describe the interaction strengths using the relative dipolar strength $\epsilon_{\rm dd}=a_{\rm dd}/a$, where $a_{\rm dd}=\mu_0\mu^2m/12\pi\hbar^2\simeq131\,a_0$ is the dipolar length, $a_0$ is the Bohr radius, $\hbar$ is the reduced Planck constant, $\mu_0$ is the vacuum permeability and $m$ is the atomic mass. 
To observe the self-bound state, we prepare an initially oblate Bose-Einstein condensate (BEC)\cite{Lu:2011} of \textsuperscript{164}Dy with an atom number of $N=6,000(500)$ at a temperature of $T=20\,$nK at large scattering length ($B_{\rm BEC}=7.089(5)$\,G), for which the interaction is contact-dominated, and shape it using an additional optical trap into a prolate shape along the magnetic field direction. 
This reshaping is done in two stages. First, we ramp up a focused beam (with a wavelength of $532$\,nm, aligned in the $z$ direction) within 50\,ms. With this attractive potential, the radial trap frequencies are increased to change the aspect ratio of the trap $\lambda=\omega_z/\omega_\rho$ from $\lambda=80\,$Hz$/20.5\,$Hz $=3.9$ down to $\lambda=80\,$Hz$/61\,$Hz $=1.3$; here, $\omega_z$ ($\omega_\rho$) is the trapping frequency along (perpendicular to) the magnetic field direction. At this point, owing to magnetostriction\cite{Lahaye:2009}, the BEC becomes prolate with a cloud aspect ratio $\kappa=\sigma_z/\sigma_\rho$ of approximately $1.5$ (with $\sigma_z$ ($\sigma_\rho$) the physical size (at 1/e$^2$) of the cloud in (perpendicular to) the field direction) and has a typical atom number estimated to be $N=3,000(300)$. Note that not all of these atoms are found to be in the self-bound state. Second, we apply a magnetic field gradient to the atomic cloud that exactly compensates the gravitational force and thus results in levitation. In this configuration, the cloud undergoes a continuous crossover from the BEC state directly to the single-droplet ground state as the scattering length is reduced, bypassing a bistable region\cite{Waechtler:2016a,Bisset:2016}. Over the next $50\,$ms we lower the field to various values between $B=6.831(5)\,-\,6.469(5)\,$G, (indicated by the hatched area in Fig. \ref{Fig:FBscheme}b) to decrease the scattering length and create a single droplet. We hold the atoms in this configuration for 10 ms before ramping the optical trap powers within $20\,$ms to approximately 5\% of their initial values, keeping a constant trap aspect ratio.
\begin{figure}[!ht]
\begin{center}
\includegraphics[width=\columnwidth]{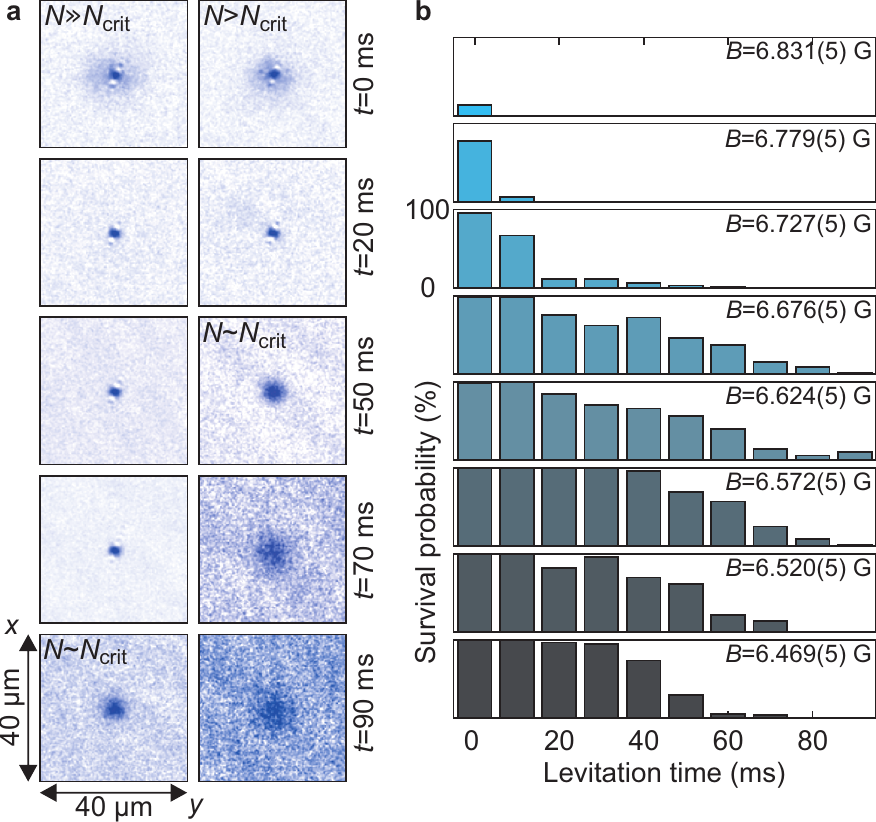}\\ 
\end{center}
\caption{\textbf{Droplet survival probability. a}, Image sequences of two droplets with different levitation times at the same magnetic field $B=6.676(5)$\,G. The images are not multiple images of the same droplet, but are selected from various images because the imaging process is destructive. All images are rescaled to the maximum optical density and have been re-centred. In the left-hand column, we start with an atom number that is much larger than the critical atom number for stable droplets and observe a single droplet up to $t_{\rm levitate}=70$\,ms. Between $t=70\,$ms and $t=90\,$ms, the cloud reaches the critical atom number and evaporates back to a gas phase, observed as an expanding cloud. In the right-hand column, the droplet starts with an atom number that is much closer to the critical atom number, leading to an earlier evaporation, between $20\,$ms and $50\,$ms of levitation time. From this point the cloud evaporates to the BEC phase and expands. \textbf{b}, Histogram of the survival probability of a single droplet as function of levitation time and magnetic field. At low scattering lengths ($B=6.469(5)$\,G), we always observe droplets for up to $t_{\rm levitate}=30$\,ms, followed by a fast decay in survival probability that is explained by a fast decay in atom number as a result of three-body collisions. For increasing scattering length, we observe an increase in the lifetime of the droplets up to a magnetic field range of $B=6.572(5)-6.676(5)$\,G. At these conditions we observe a single droplet with a size below our resolution after a levitation time of $t_{\rm levitate}=90$\,ms. Further increase of the scattering length shows a fast decay of self-bound droplets already for short times ($t_{\rm levitate}=20$\,ms), which we interpret as originating from an increase in the critical atom number to values close to our initial atom number. For the highest scattering length ($B=6.831(5)$\,G), we barely create droplets in the trap.}
\label{Fig:Pictures}
\end{figure}
At this point, we suddenly turn off the trap and image the cloud using far-detuned phase-contrast polarization imaging after various levitation times up to $t_{\rm levitate}=90$\,ms. This sequence is shown schematically in Fig.\ref{Fig:FBscheme}a. Being sensitive only to high densities, we observe that a thermal fraction of the atomic cloud expands very quickly, whereas a very small and dense cloud remains for very long times. We interpret this observation as a self-bound quantum droplet. We calculated the radial size of the quantum droplets to be approximately 300\,nm, which is smaller than our imaging resolution of $1\,\mu$m such that we observe astigmatic diffraction (see Fig. \ref{Fig:Pictures}a)). At specific fields, we observe these droplets for times as long as $t_{\rm levitate}=90\,$ms. At some time during the trap-free levitation, we observe that the droplets have expanded. We reason that this behaviour is due to the fact that droplets lose atoms as a result of three-body decay or residual excitations until they reach a critical atom number, below which they are no longer self-bound and evaporate back into a gas phase. 
\begin{figure}[!htbp]
\begin{center}
\includegraphics[width=0.99\columnwidth]{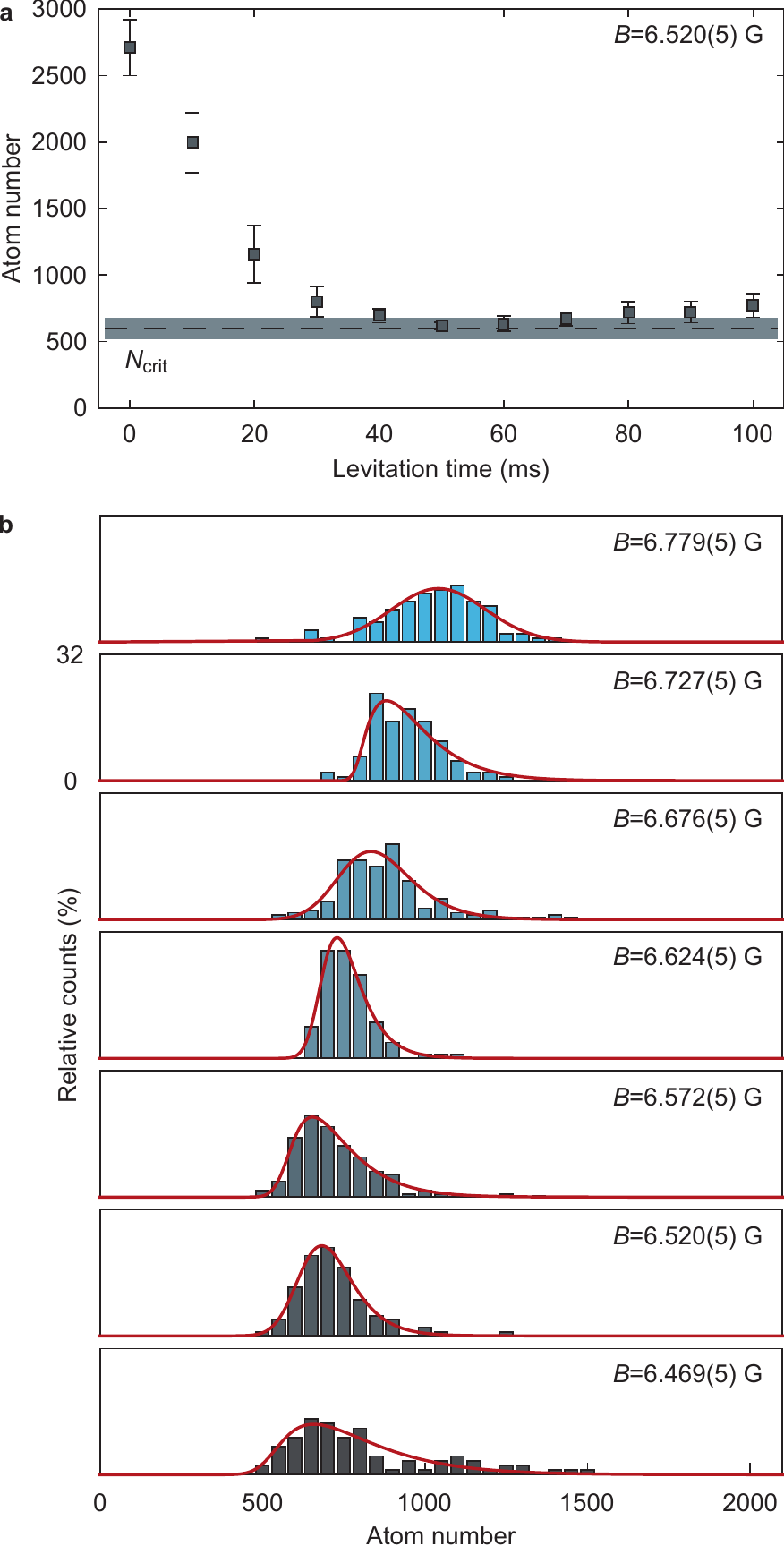}\\
\end{center}
\caption{\textbf{Critical atom number. a}, Decay in atom number as function of levitation time. Each point represents the mean atom number of 20 realizations; error bars denote the standard deviation. We observe a decay for short times to an essentially constant number for long times. The dashed line shows the critical atom number as determined by the best fit of our convoluted model (see text) to the data; the shaded area shows the error in this fit. 
\textbf{b}, We analyse the atom number distribution for levitation times in the range $t_{\rm levitate}=60-100$\,ms because the atom number is mostly constant in this range. We bin the atom number to a window of 50 atoms and plot the relative counts as a function of atom number and magnetic field. The red curves represent fits of the convoluted functions to the observed histograms. The colours of the plotted histograms match those in Fig. \ref{Fig:Pictures}b, and represent the magnetic field change.}
\label{Fig:Hist}
\end{figure}
In this context, the word 'evaporation' is used to denote the transition from a dilute self-bound liquid state to an expanding gas state. Given our shot-to-shot noise in the initial atom number, the critical atom number is reached at different times. This behaviour is represented in Fig. \ref{Fig:Pictures}a). 

As a first analysis, we count the images in which we observe a single droplet over 100 realizations and plot the survival probability for different magnetic fields (Fig.\ref{Fig:Pictures}b). The levitation time is varied between $t_{\rm levitate}=0$\,ms, which essentially represents a trapped cloud, and $t_{\rm levitate}=90$\,ms. For low scattering length ($B=6.469(5)$\,G), we always create a single droplet, but its lifetime is short. As the scattering length increases, so does the lifetime. We find a maximal survival probability in the magnetic field range $B=6.572(5)-6.676(5)$G. For even higher scattering lengths, we find droplets only at $0\,$ms, and very few self-bound droplets. The calculated survival probabilities are in qualitative agreement with an increasing critical atom number and a decreasing rate of atom loss in the droplets with increasing scattering length. This behaviour has been observed\cite{Ferrier:2016} in a waveguide configuration and for a single trapped droplet\cite{Chomaz:2016}, and is supported by calculations on a self-bound droplet\cite{Waechtler:2016b,Baillie:2016}. However, the precise evolution depends on the spread in initial atom number and the fact that droplets evaporate at different atom numbers (see below).
\begin{figure}[t]
\includegraphics[width=0.99\columnwidth]{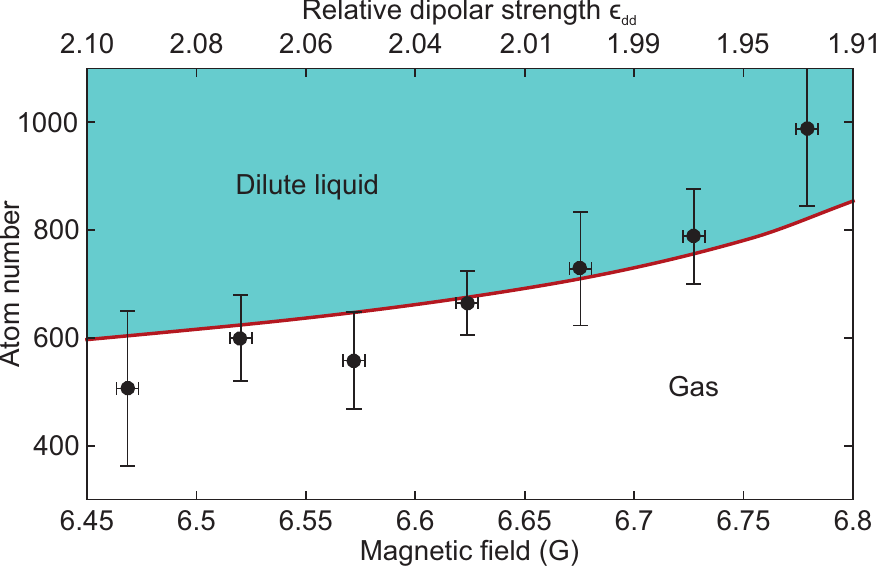}\\
\caption{\textbf{Phase transition between dilute liquid and gas.} The data points show the critical atom number as a function of the magnetic field, as determined from the fit values from Fig. \ref{Fig:Hist}b. The error in the atom number is given by the quadratic mean of the widths of the Gaussian and Maxwell-Boltzmann distributions; the error in magnetic field describes the resolution of our magnetic field coils. As the magnetic field decreases, so does the critical atom number $N_{\rm crit}$. We identify the upper-left corner as the dilute liquid phase and the lower-right corner as the gas phase. The critical atom number describes the phase transition between dilute liquid and gas. The solid red line represents a full Gross-Pitaevskii simulation for different relative dipolar strengths $\edd$.}
\label{Fig:CritN}
\end{figure}

To obtain a more quantitative analysis of the critical atom number of these droplets, we intentionally evaporate them after variable levitation times by increasing the magnetic field to $B_{evap}=6.986(5)$\,G (dashed black line in Fig. \ref{Fig:FBscheme}b). At this field, we observe that all droplets have been evaporated and interpret this to mean that the critical atom number at this field is higher than all relevant atom numbers observed here. After expansion, the atom number can be determined accurately without being limited by the finite resolution of the imaging optics. Here we observe that the number of atoms in the droplets decays to an essentially constant number - further indication of a critical atom number for self-bound droplets. This behaviour is demonstrated in Fig. \ref{Fig:Hist}a for a magnetic field of $B=6.520(5)$\,G, wherein each point is represented by a mean atom number that is calculated from 20 images, and the error denotes one standard deviation. A histogram of the atom number distribution for long levitation times ($t_{\rm levitate}\geq60$\,ms) and for different magnetic fields is shown in Fig. \ref{Fig:Hist}b. We observe that the atom number distributions shift with scattering length, and conclude that the droplets lose atoms until they reach the critical atom number, at which point all of the atoms evaporate out of the droplets into the gas phase. We observe that at long times, when most droplets have evaporated, there is an asymmetric dispersion in atom number to higher values. We posit that this reflects the fact that not all droplets evaporate at exactly the critical atom number, but that the evaporation for some droplets can occur with $N>N_{\rm crit}$, owing to the presence of residual (for example, thermal) excitations in the droplet. To extract a critical atom number we fit the histograms with a convolution of a Gaussian and a Maxwell-Boltzmann distribution (see Methods).

The best fits are shown in Fig. \ref{Fig:Hist}b as red curves. We plot our result of the critical atom number in Fig. \ref{Fig:CritN} and compare it to full, extended Gross-Pitaevskii simulations (see Methods). The error is given by the quadratic mean of the widths of the Maxwell-Boltzmann and Gaussian distributions. 
This way of determining $N_{\rm crit}$ depends on the model used to determine the fit; other definitions could lead to slightly different values. Nevertheless, we see a clear change in the critical atom number with magnetic field, and with this we probe the phase transition line between the dilute liquid phase and the gas phase. To compare the results with the simulations, we calculate the relative dipolar strength for our magnetic field range. To do so, we include the Feshbach resonance at $B_{01}=7.117(3)$\,G with a width of $\Delta B_1=51(15)$\,mG and a resonance at $B_{02}=5.1(1)$\,G with a width of $\Delta B_2=0.1(1)$\,G. A best fit is obtained when we change the previously assumed local background scattering length\cite{Tang:2015} from $a_{\rm bg}=92(8)\,a_0$ to $a_{\rm bg}=62.5\,a_0$. This lower value seems, at first, to be incompatible with previous measurements at different fields\cite{Tang:2015}; however, the complexity of the scattering problem in dysprosium does not allow a theoretical prediction and the local $a_{bg}$ might vary in other ranges of magnetic field. In addition, theoretical simulations of the Rosensweig instability\cite{Waechtler:2016b,Bisset:2016} suggest that a background scattering length of less than $92\,a_0$ is necessary to agree with experimentally observed timescales\cite{Kadau:2016}. In our measurements, the strong dependence of $N_{\rm crit}$ on scattering length provides a very high sensitivity. Changing the background scattering length from $92\,a_0$ to $62.5\,a_0$ reduces $N_{\rm crit}$ by almost a factor of ten. This method therefore enables a very precise measurement of the scattering length. However, at this level of precision, we must question the approximations made in our model, such as the first-order Born approximation for the dipolar scattering and the local density approximation. Consequently, the value of $a_{\rm bg}$ quoted here is model-dependent, and could be subject to future corrections. An independent measurement of $a$ via the methods of ref. 12 for instance, would make $N_{\rm crit}$ measurements a very sensitive benchmark for many-body theories.

By removing the need for any trapping potential, our observation of the self-bound regime offers access to truly isolated, dissipative quantum systems in which the effective cancellation of the mean field enables quantum correlations to be studied in detail. The gas-to-liquid transition and, in particular, the nucleation dynamics of the droplets will be sensitive probes of the interplay between interactions and quantum correlations.

\begin{methods}
\noindent\textbf{Convolution model} To extract the critical atom number from the data in figure \ref{Fig:Hist} we fit the histograms with a phenomenological model (represented as red lines). This model consists of the convolution of a Gaussian and a Maxwell-Boltzmann distribution. The symmetric Gaussian distribution represents broadening effects that result from statistical errors including detection noise. The asymmetric Maxwell-Boltzmann distribution is used to model the possibility of a droplet fully evaporating at atom numbers higher than the critical atom number, as a result of the presence of collective excitations in the droplets. From the fit we extract the critical atom number and two widths, one from each distribution in the convolution. We represent the quadratic mean of these widths as error bars in Fig. \ref{Fig:CritN}.

\noindent\textbf{Extended Gross-Pitaevskii simulation} To compare our results to current theory\cite{Waechtler:2016a,Baillie:2016}, we perform simulations of the effective Gross-Pitaevskii equation 
\begin{eqnarray}
\begin{aligned}\label{eq:GPE}
i\hbar \partial_t \psi &= \bigg[ -\frac{\hbar^2\nabla^2}{2 m} + V_{ext}	+ g |\psi|^2 \\
&+ \int\!\mathrm{d}\mathbf{r}^\prime V_{dd}(\mathbf{r}-\mathbf{r}^\prime) |\psi(\mathbf{r}^\prime)|^2 \\ 
&+ \frac{32 g \sqrt{a^3}}{3 \sqrt\pi} (1 + \frac{3}{2} \varepsilon_{dd}^2) |\psi|^3 -i \frac{\hbar}{2} L_3 |\psi|^4 \bigg] \psi(\mathbf{r})
\end{aligned}
\end{eqnarray}
using a simple interaction potential, and taking into account quantum fluctuations within a local density approximation\cite{Lima:2011,Lima:2012} and three-body losses. Here
\begin{equation}
V_{dd}(\mathbf{r})=\frac{\mu_0\mu^2}{4\pi}\frac{1-3\cos^2(\vartheta)}{|\mathbf{r}|^3}
\label{eq:Vdd}
\end{equation}
describes the dipole-dipole interaction potential, with $\vartheta$ denoting the angle between the polarization direction of the dipoles and their relative orientation. The main assumptions of this model are therefore the validity of the local density approximation and of the interaction potential, which results from the first-order Born approximation. The magnetic moment and scattering length are $\mu=9.93\,\mu_\mathrm{B}$ and $a = 60 - 80 \,a_0$, respectively. The latter defines $g = 4 \pi a \hbar^2 / m$ and is chosen such that we are in agreement with the critical atom numbers we observe in the experiment. The loss parameter $L_3 = 1.25\cdot10^{-41}\,\mathrm{m^6/s}$ is estimated from measurements on a thermal cloud and is assumed to be constant over the small range of scattering lengths. The validity of the local density approximation is supported by quantum Monte Carlo simulations\cite{Saito:2016} and recent measurements with erbium atoms\cite{Chomaz:2016}.\\
To obtain the data in figure \ref{Fig:CritN} we choose $V_{ext} = 0$ and initially prepare $N_0 > N_{\rm crit}$ atoms with a gaussian density distribution ($\sigma_r = 250\,\mathrm{nm}$, $\sigma_z = 1500\,\mathrm{nm}$). The ground state is reached by imaginary time evolution of equation (\ref{eq:GPE}) using a splitstep Fourier method. Following this preparation of the self-bound droplet with $N_0$ atoms, we simulate the dynamics via real-time evolution. Because the atom number $N < N_0$ decays, owing to three-body losses, the density and the effective two-body attraction are also reduced. At $N = N_{\rm crit}$, the contributions by the effective two-body attraction and the quantum pressure are the same in magnitude, and the droplet evaporates quickly. This evaporation process manifests itself as a decrease in peak density of at least one order of magnitude. Three-body losses are highly suppressed then, such that the atom number stays almost constant for an evaporated droplet.
\end{methods}

\bibliographystyle{naturemag}
\bibliography{QuantumLiquid}

\begin{addendum}
 \item[Acknowledgements] We thank H. P. B\"uchler, L. Santos, F. Ferlaino, W. Ketterle, H. Sadeghpour, M. Zwierlein and V. Vuleti\'{c} for discussions. This work is supported by the German Research Foundation (DFG) within SFB/TRR21 as well as FOR 2247. I.F.B aknowledges support from the EU within Horizon2020 Marie Sk\l odowska Curie IF (703419 DipInQuantum).
\end{addendum}


%
%
%

\end{document}